\documentclass[twocolumn]{aastex61}

\newcommand{\pd}[2]{\frac{\partial #1}{\partial #2}}

\renewcommand{\vec}[1]{\mathbf{#1}}

\usepackage{amsmath}

\submitjournal{ApJ}

\shorttitle{Radio Evolution of Supernova Remnants}
\shortauthors{Pavlovi\'c et al.}

\begin{document}

\title{Radio Evolution of Supernova Remnants Including Non-Linear Particle Acceleration: Insights from Hydrodynamic Simulations}


\correspondingauthor{Marko Z. Pavlovi\'c}
\email{marko@matf.bg.ac.rs}

\author[0000-0001-7633-8110]{Marko Z. Pavlovi\'c}
\affiliation{Department of Astronomy \\
Faculty of Mathematics \\
University of Belgrade
Studentski trg 16, 11000 Belgrade, Serbia}

\author[0000-0003-0665-0939]{Dejan Uro\v sevi\'c}
\affiliation{Department of Astronomy \\
Faculty of Mathematics \\
University of Belgrade
Studentski trg 16, 11000 Belgrade, Serbia}
\affiliation{Isaac Newton Institute of Chile\\
Yugoslavia Branch}

\author[0000-0002-8036-4132]{Bojan Arbutina}
\affiliation{Department of Astronomy \\
Faculty of Mathematics \\
University of Belgrade
Studentski trg 16, 11000 Belgrade, Serbia}

\author[0000-0003-2836-540X]{Salvatore Orlando}
\affiliation{INAF -- Osservatorio Astronomico di Palermo 
\lq\lq G.S. Vaiana\rq \rq\\
Piazza del Parlamento 1, I-90134 Palermo, Italy}

\author[0000-0003-2762-8378]{Nigel Maxted}
\affiliation{The School of Physics \\
The University of New South Wales, Sydney, 2052, Australia}
\affiliation{Western Sydney University \\
Locked Bag 1797, Penrith South DC, NSW 1797, Australia}

\author[0000-0002-4990-9288]{Miroslav D. Filipovi\' c}
\affiliation{Western Sydney University \\
Locked Bag 1797, Penrith South DC, NSW 1797, Australia}

\begin{abstract}

We present a model for the radio evolution of supernova remnants (SNRs) obtained by using three-dimensional (3D) hydrodynamic simulations, coupled with nonlinear kinetic theory of cosmic ray (CR) acceleration in SNRs. We model the radio evolution of SNRs on a global level, by performing simulations for wide range of the relevant physical parameters, such as the ambient density, the supernova (SN) explosion energy, the acceleration efficiency and the magnetic field amplification (MFA) efficiency. We attribute the observed spread of radio surface brightnesses for corresponding SNR diameters to the spread of these parameters. In addition to our simulations of type Ia SNRs, we also considered SNR radio evolution in denser, nonuniform circumstellar environment, modified by the progenitor star wind. These simulations start with the mass of the ejecta substantially higher than in the case of a type Ia SN and presumably lower shock speed. The magnetic field is understandably seen as very important for the radio evolution of SNRs. In terms of MFA, we include both resonant and non-resonant modes in our large scale simulations, by implementing models obtained from first-principles, particle-in-cell (PIC) simulations and non-linear magnetohydrodynamical (MHD) simulations. We test the quality and reliability of our models on a  sample consisting of Galactic and extragalactic SNRs. Our simulations give $\Sigma-D$ slopes between -4 and -6 for the full Sedov regime. Recent empirical slopes obtained for the Galactic samples are around -5, while for the extragalactic samples are around -4.


\end{abstract}




\section{Introduction} 
\label{sec:intro}

We expect for future radio observations to bring important advances in understanding the properties of the many high-energy sources, including supernova remnants (SNRs). Putting into operation some of the new generations of radio telescopes will inevitably lead to the detection of many new SNRs, possibly alleviating the incompleteness of the current Galactic and extragalactic SNR samples. In order to take full advantage of these new observations, we must fully understand the radio evolution of SNRs, the intrinsic and environmental diversity of SNRs, their evolutionary status and implications for cosmic ray acceleration, the supernovae (SNe) rate and origin as well as the energy input into the ISM.

Today, it is widely accepted that cosmic rays (CRs) are accelerated up to, and possibly beyond $10^{15}$ eV (known as the 'knee') at the shock waves
of SNRs. The most efficient mechanism for accelerating high energy CRs is diffusive shock acceleration (DSA) proposed by \citet{krym77}, \citet{axford77}, \citet{bell78a, bell78b} and \citet{blandford78}, providing the energy gain due to multiple collisions with irregularities of the magnetic field. During the past decades, effort has been made to develop the extension of DSA to the case in which CRs are not simply test-particles but also influence the shock dynamics \citep[][]{caprioli12, blasi13}. Non-linear theories of DSA (known as NLDSA) predicts the back-reaction of the accelerated CRs to induce in the upstream the formation of a so-called precursor, supported by recent observational evidence \citep{knez17}.

High resolution mapping of the Balmer dominated shocks in SN 1006  suggests the presence of suprathermal protons as potential seeds of high-energy CRs \citep{nikolic13}. However, unambiguous evidence of CR hadron acceleration in supernova remnants exists only for a few sources \citep[e.g. Tycho, W44, IC443, Vela Jr.,][]{morlino12, acker13, fakui17}. On the other hand, highly energetic electrons efficiently emit radiation from the radio to the X-ray band through the synchrotron (magneto-bremsstrahlung) mechanism and their detection is far easier.

Radio emission has been detected for more than half a century, still remaining the most common diagnostic tool for SNRs and a cornerstone in this field. The large majority of all known SNRs are sources of radio-synchrotron emission, testifying the non-thermal processes ongoing there due to the existence of relativistic electrons. There are several very young SNRs (up to a few hundred years old) which are attractive 'laboratories', allowing us to study radio evolution almost from the very beginnings. Fortunately, we have a considerable amount of multiwavebands observations for them. We can test our models for these objects and then apply it to a broader sample of SNRs in the Galaxy and even further. SN 1987A has enabled the observation of a peculiar class of Type II events at close proximity \citep{zanardo10, call16}. Since the detection, the intensity of the SNR radio emission has shown a steady increase, surpassing the initial radio brightness. 
The SNR originating from this explosion can be even used as a template to link SNe to their remnants \citep{orlando15}. The youngest known Galactic SNR, G1.9+0.3 also provides unique information about the particle acceleration and broad-band emission at the early stages of evolution of SNRs \citep{green08, murphy08, horta14, aharonian17}. 
On the other hand, the brightest extrasolar radio source in the sky, SNR Cassiopeia A, shows the opposite trend. The synchrotron flux density in radio has been decreasing at a rate of 0.6$-$0.7\% year$^{-1}$ at 1 GHz \citep{baars77, raich00}. \citet{handbook16} attribute this flux decrease is because of adiabatic and radiative losses of relativistic particles with expansion, but the details might depend  on particle acceleration processes as well as the physical structures of SN ejecta and the surrounding medium.

\citet{sklovski60a} initially predicted variation in the radio flux density of the SNR Cassiopeia A, attributing it to the expansion of the remnant and the associated decrease in its magnetic field. He established the so-called radio surface-brightness-to-diameter ($\Sigma-D$) relation for SNRs, representing the radio evolutionary path, and also proposed its usage as an SNR distance determination method \citep{sklovski60b}. Modelling such a complex phenomena without taking into account widespread intrinsic properties of individual SNRs inevitably leads to a large scatter in the observed $\Sigma-D$ distribution of SNRs. The combined effect of evolutionary tracks of objects with different initial explosion energies, mass of ejected matter, magnetic field strength, in very different ambient conditions etc, together with selection effects \citep[e.g.][]{green91, urosevic05, urosevic10}, requires caution when using the relation as a distance estimator. 

Leaving aside the physical flaws, biases and selection effects, care also has to be taken to apply appropriate statistical treatment of the SNR radio evolution and significant progress has been made in recent years \citep{vukotic14}.

\citet{ferrand12, ferrand14} and \citet{orlando12, orlando15, orlando16} clearly demonstrated the full potential of high-resolution, 3D simulations in SNR evolution studies, reproducing the main observables of the SNRs and the properties of their broad-band emission. The development of hydrodynamic instabilities at the contact discontinuity can be modelled numerically in 3D to allow an accurate description of the downstream plasma structure, particularly in the mixing region between the forward and reverse shocks. Studies of radio emission will benefit the most from this type of modeling  because the radio continuum emission mainly originates from this region.

The modeling presented in our paper should provide a framework
for the interpretation of current SNR radio observations, as well as for 
the preparation of observations with future radio instruments, in particular ALMA\footnote{The Atacama Large Millimeter/submillimeter Array},
MWA\footnote{The Murchison Widefield Array}, ASKAP\footnote{The Australian Square Kilometre Array Pathfinder}, SKA\footnote{The Square Kilometre Array} and FAST\footnote{The Five-hundred-meter Aperture Spherical radio Telescope, the largest and most sensitive single dish radio telescope in the world}.


\section{The Model and numerical setup} 
\label{sec:model}

\subsection{Modeling the Dynamical Evolution of an SNR}
\label{subsec:dynamics}

We modelled the dynamical evolution of SNRs by numerically solving the time-dependent Euler partial differential equations (PDEs) of fluid dynamics, also known as hyperbolic conservation laws, that we write as:

\begin{equation}\label{eq:CL}
  \pd{\vec{U}}{t} + \nabla\cdot\vec{F} = 0.
\end{equation}

\noindent Here $\vec{U}$ and $\vec{F}$ represent a state and flux vectors, respectively, 
which can be written in the form:

\begin{eqnarray}
\vec{U}  & = &
 \left(
\begin{array}{c}
\rho  \\
\rho \boldsymbol{\bf \upsilon}\\
\rho E \end{array} \right),~
\vec{F}= \left(
\begin{array}{c}
\rho  \boldsymbol{\bf \upsilon}\\
\rho  \boldsymbol{\bf \upsilon} \boldsymbol{\bf \upsilon}^{T} + P  \\
(\rho E + P) \boldsymbol{\bf \upsilon} \\
\end{array}     \right),
\end{eqnarray}


\noindent where $E = \epsilon + |\boldsymbol{\upsilon}|^2/2$ is the total gas energy per unit mass (sum of the internal energy $\epsilon$, and kinetic energy),
$\rho = \mu m_H n_{\rm H}$ is the mass density, $\mu=1.4$ is the mean
atomic mass (assuming cosmic abundances, namely a 10:1 ratio for H:He), $m_{\rm H}$ is the mass of the hydrogen atom, $n_{\rm H}$ is the hydrogen number density and $\boldsymbol{\upsilon}$ is the gas velocity vector. As a thermodynamic closure condition for the system, we used the ideal gas equation of state (EoS), $P=(\gamma-1) \rho \epsilon$, where $\gamma$ is the adiabatic index. We performed 3D simulations in Cartesian geometry ($x,y,z$), neglecting radiative losses and thermal conduction.

Our simulations are performed by using the publicly available, Godunov-type code for astrophysical gasdynamics PLUTO \citep[Version 4.2;][]{mignone07, mignone12}. To overcome the spatial and temporal scale challenges in the problems considered, we
rely on the block-structured, adaptive mesh refinement (AMR) implementation in the PLUTO code, based on the Chombo library\footnote{https://commons.lbl.gov/display/chombo}. The code uses a distributed infrastructure for parallel computations through the message passing interface (MPI) standard. We used the following set of PLUTO standard algorithms: {\sf linear} interpolation with default limiter, {\sf HLLC} Riemann solver, {\sf RK2} for the time evolution and {\sf MULTID} flattening for the numerical dissipation near the strong shocks. We employ 9 nested levels of resolution, with resolution increasing two-fold at each refinement level, placed on a base (non-refined) grid of $32^3$, leading to maximum AMR resolution of $16384^3$ (which is used for simulations where maximum size of physical grid is equal or exceeds 80 pc, see Table~\ref{tab:Param}). In order to lower computational costs and keep them approximately constant as the blast wave expands, the maximum number of refinement levels decreases from 9 (initial ejecta profile) to 3 (at the end of evolution), as suggested and previously implemented by \citet{orlando12} in FLASH code. We record the shock position during the entire SNR evolution and consequently calculate the required number of refinement levels for a particular time, which is then forwarded to Chombo library interface.

As initial conditions for the SN ejecta, we adopt the exponential density and velocity profiles of a post-deflagration stellar remnant as proposed by \citet{dwarka98}. They showed that the exponential density profile gives the best approximate representation in comparison with the power-law and constant ejecta density cases. This type of ejecta profile is adopted for all modeled SNRs, whether it originates from type Ia (thermonuclear) or core-collapse (CC) SN although, this may not be completely adequate in the latter case. The radial profiles of the ejecta density should not significantly affect the radio emission, especially at later times. As pointed out by \citet{dwarka98}, the exponential profile predicts a density
curve increasing from the reverse shock to the contact discontinuity,
while the power law profile gives a decreasing density in the same direction. This may affect radio morphology during the earliest stages only, as well as the clumpiness of the ejecta \citep{orlando12}. We assumed total ejecta mass equal to the Chandrasekhar mass $M_{\rm ej} = 1.4M_{\sun}$ for type Ia and higher ejecta mass $M_{\rm ej} = 10M_{\sun}$ for CC explosions. Note that we assume here that type Ia SNe are the result of a thermonuclear runaway reaction triggered by accretion onto a C/O white dwarf (WD) from a non-degenerate companion star. However, if we consider an explosion triggered by the merger of two WDs in a compact binary system, as suggested by growing evidence for a small sample of SNe \citep{gilf10, olling15, maggi16, woods17}, the total mass and energy could be considerably higher and may affect the dynamics and radio emission.

We restrict our simulations to the case of an isotropic, warm interstellar medium (ISM) of temperature $T=10^4~$K. Simulations follow SNR evolution for five ISM phases, with hydrogen number densities $n_{\rm H}=0.005, 0.02, 0.2, 0.5$ and 2 cm$^{-3}$. These values roughly cover typical estimates for ambient densities of individual Galactic \citep{arbutina05} and extragalactic SNRs \citep{berkhujsen86}. The constant density approximation is not expected to influence the total radio emission, but we note that inhomogeneity becomes important in morphological studies (see \citet{slavin17} for the basic ideas about simulations of SNRs in cloudy medium, although they are mainly interested in consequences for the X-ray emission; and \citet{kostic16} for the influence of fractal density structure of the ISM on the radio evolution for SNRs).

We simulate SNRs originating from explosions with initial total energies $E_{0}=(0.5, 1.0~ \rm{and}~ 2.0) \times 10^{51}$ erg. We assume in our 3D simulations that almost all ($>$ 98\%) of the explosion energy is kinetic. This is shown by \citet{orlando16} to be valid assumption very early in the evolution of an SNR, even a few days after the SN explosion. The flow becomes homologous soon after the SN explodes and therefore, velocity increases linearly with distance from center to the outer edge of the ejecta where it reaches a maximum value, $V_{\rm ej}^{\rm max}$. Although SNe eject a mass of material with a range of velocities, the characteristic
initial explosion velocity is of order $\sim 10^4~\rm{km}~\rm{s}^{-1}$ for a Type Ia and $\sim 5000~\rm{km}~\rm{s}^{-1}$ for a CC event \citep{reyn08}.
For type Ia SNRs, we adopt  $V_{\rm ej}^{\rm max}=$ 20\,000\ km s$^{-1}$ for referent cases with initial total energy $E_{0}=1.0\times 10^{51}$ erg and adjust it for lower or higher explosion energies (without changes in density profile). For CC SNRs, it is reasonable to assume lower ejecta velocities, namely $V_{\rm ej}^{\rm max}=$10\,000 km s$^{-1}$ for the most energetic explosions.

Simulations that assume ISM phases with high ambient densities, namely $n_{\rm H}=0.5$ and $2~$cm$^{-3}$, are expected to primarily represent SNRs that originate from the collapse of the cores of massive progenitor stars (CC; belonging to Type II, Type Ib and Type Ic). What is important, at least for the early interaction between the SN and ambient medium, is the mass-loss immediately before the explosion. These massive progenitor stars have slow winds with typical velocities 10$-$50 km s$^{-1}$ and the 
mass-loss rates in the range 10$^{-6}$$-$10$^{-5}M_{\odot}$ yr$^{-1}$ \citep[see, for example,][]{reyn17}. With the typical assumption that the gas density in the wind is proportional to $r^{-2}$ (where $r$ is the radial distance from the center of the explosion), we model the density profile encountered by the SNR shock with:

\begin{equation}
n(r) = n_{\rm{H}} + n_{\rm{w}} = n_{\rm{H}} + \frac{\dot{M}_{\rm w}}{4\pi r^2 \upsilon_{\rm w} \mu m_{\rm H}}
\label{wind_dens}
\end{equation}

\noindent where we assumed a spherically symmetric wind with a mass-loss rate of $\dot{M}_{\rm w}=10^{-5}M_{\odot}~\rm{yr}^{-1}$ and wind velocity $\upsilon_{\rm w} = 10~$km s$^{-1}$. It is likely that an interaction region/layer with increased density exists between the wind and the surrounding ISM, but we have neglected this in this initial study. However, we will address this in our subsequent study.
Also note that, in Equation \ref{wind_dens}, we add an isotropic wind
component on top of the constant density, therefore overestimating the density at any one point by the amount of the constant offset. We do not expect a significant effect on the blast wave dynamics, taking into account time and spatial scales of our simulations. For a detailed study of SNR interactions with the circumstellar medium (CSM) see \citet{orlando15}. 

We neglect radiative losses and therefore run our simulations only while the adiabatic condition is completely applicable. The transition from fully adiabatic to fully radiative shock is not very sharp and lasts for almost equal time as the adiabatic stage, representing the so-called 
"post-adiabatic" phase \citep{petruk16}. The full adiabatic regime ends at around the transition time (the earliest cooling of any fluid parcel) and this marks the beginning of the post-adiabatic phase \citep{blondin98}:

\begin{equation}
t_{\rm tr} = 2.83\times 10^4 E_{51}^{4/17} n_{\rm H}^{-9/17}~\mbox{yr}~,
\label{trans_time}
\end{equation}

\noindent while this phase changes to the radiative phase around the
shell-formation time $t_{\rm sf} \approx 1.8t_{\rm tr}$ 
\citep{cox82,petruk16}, where $E_{51} = E_0/(10^{51} \rm{erg})$. At transition time, a Sedov-Taylor (ST) blast wave has reached a radius of $R_{\rm ST} \approx 19.1E_{51}^{5/17} n_{\rm H}^{-7/17}~\mbox{pc}$ and the velocity of a ST blast wave at this age
is $V_{\rm ST} \approx 260 E_{51}^{1/17} n_{\rm H}^{2/17}~\mbox{km/s}$ \citep{blondin98}. It is reasonable to assume that radio SNRs are observed approximately until the end of the Sedov phase, when their radio emission decreases significantly \citep{bandiera10}. \citet{bandiera10} and \citet{bozz17} even provide, from their statistical study, a good argument for SNRs being mostly visible around the end of the adiabatic stage. Because of this, and also due to neglected radiative cooling effects, in the entire set of our simulations, we follow the hydrodynamic and radio evolution strictly before reaching the transition time.

We assume initially spherical remnants with radius $R_0$ = 0.5 pc
(initial age of $\approx$ 30 yr) beginning its evolution from the origin of the 3D Cartesian coordinate system $(x_0, y_0, z_0)=(0, 0, 0)$ and we only simulate one octant of the SNR. Our computational domain extends from 20 to 200 parsecs in the $x, y$ and $z$ directions, depending on the transition time (and its corresponding final radius) for a particular SNR. We assume zero-gradient (outflow) boundary conditions at all boundaries.
For the simulations of core-collapse SNRs (evolving in denser ISM), chosen parameters and Equation~\ref{wind_dens} give stellar wind density at initial radius 0.5 pc of $n_{\rm{w}}\approx$~9 cm$^{-3}$.

\subsection{Non-linear diffusive shock acceleration}
\label{subsec:nldsa}

We perform our 3D hydrodynamic modelling by including back-reaction of
accelerated CRs and consistent treatment of magnetic field amplification (MFA), as previously done in \citet{pavlovic17} (hereafter referred to as P17).

\citet{pformer17} developed new methods to integrate the CR evolution equations coupled to MHD on an unstructured moving mesh, implemented through the AREPO code, mainly intended for cosmological simulations.
AREPO follows advective CR transport within the magnetized plasma, as well as anisotropic diffusive transport of CRs along the local magnetic field. They showed that CR acceleration at blast waves does not significantly break the self-similarity of the ST solution and that the resulting modifications can be approximated by a suitably adjusted adiabatic index, as done in our approach.

Detection and tracking of SNR shock waves in the fluid, travelling in some direction $x$, is based on two standard numerical conditions, namely $\nabla\boldsymbol{\upsilon} < 0$ and $\Delta x \frac{\nabla P}{P} > \varepsilon_{\rm p}$, where $\varepsilon_{\rm p}$ determines the shock strength. In the block-structured AMR approach, the cells which require additional resolution are covered with a set of rectangular grids characterized by a finer mesh spacing \citep{mignone12}. Shock detection is therefore, slightly modified in comparison with P17 because we have to pay special attention to the particular mesh level used for shock detection. In order to achieve the highest accuracy, shock detection is applied in the finest mesh level. Refinement criteria, used in our simulation, assures that zones around forward shock are tagged for maximum refinement.

We modified AMR PLUTO modules in order to couple the hydrodynamical evolution of the remnant with particle acceleration. We adopted hydrodynamic equations to use the space and time-dependent adiabatic index $\gamma_{\rm eff} = \gamma_{\rm eff}(x,y,z,t)$ i. e. $P = (\gamma_{\rm eff}-1) \epsilon$ \citep{ellison04}. The effective adiabatic index $\gamma_{\rm{eff}}$ produces the same total compression $R_{\rm tot}$ as obtained from a non-linear model \citep{blasi04, blasi05}. It is calculated at the shock front and then advected within the remnant. To fulfill this requirement, the adiabatic index (time dependent at the shock front) was treated as PLUTO built-in code feature called \lq passive scalar\rq~(or \lq colour\rq), denoted by $Q_{k}$, obeying the simple advection equation of the form:

\begin{equation}
\frac{{\rm{D}} Q_{k}}{{\rm{D}} t} = 0,
\label{postadiab:tracer}
\end{equation}

\noindent where $\frac{\rm{D}}{\rm{D} t} = \frac{\partial}{\partial t} + 
\boldsymbol{\upsilon} \cdot \nabla$ denotes the Lagrangian time derivative.
Effective adiabatic index, mimicking the presence of CRs in gas, is advected over the entire hierarchy of levels of refinement.

\subsection{Magnetic field amplification}
\label{subsec:mfa}

We slightly improved the treatment of magnetic field amplification (MFA)
in comparisson to P17, where we assumed that resonant streaming instabilities are the dominant factor for MFA. Throughout the SNR evolution, two different types of streaming instabilities are responsible for MFA \citep{amato09}. \citet{amato09} showed that the non-resonant modes are relevant mostly in the free expansion and early Sedov-Taylor phase, while resonant waves dominate in later stages of SNR evolution. \citet{bykov14} were among the first to include turbulence growth from the resonant CR streaming instability together with the non-resonant (short- and long-wavelength) CR-current-driven instabilities, in their nonlinear Monte Carlo model of efficient DSA. \citet{sarb17} also considered both contributions to ensure better theoretical background for their statistical analysis. If non-resonant modes dominate, the amplified magnetic field saturates to a value $B^2/8\pi \sim \frac{1}{2} \frac{\upsilon_{\rm s}}{c} \sigma_{\rm cr} \rho_0 \upsilon_{\rm s}^2$ \citep{bell04}, while \citet{capri14} showed that $B^2/B_0^2 \approx 3 \sigma_{\rm cr} \tilde{M_{\rm A}}$ is valid for MFA with a significant contribution from resonant modes. Here, $B$ represents the amplified field, $B_0$ the ambient magnetic field strength, $\upsilon_{\rm s}$ the shock velocity, $\sigma_{\rm cr}$ is the CR pressure\footnote{For upstream particles, with distribution $f_0$ ranging from momenta $p_{\rm min}$ to $p_{\rm max}$, their pressure $P_{\rm cr,0}$ can be computed as  $P_{\rm cr,0}=\int_{p_{\rm min}}^{p_{\rm max}} \frac{p\upsilon}{3} f_0(p)4\pi p^2 dp$, where $\upsilon(p)$ is the velocity of a particle of momentum $p$.} at the shock normalized to $\rho_0 \upsilon_{\rm s}^2$, the ambient medium density 
$\rho_0$ and $\tilde{M_{\rm A}}=(1+1/R_{\rm tot})M_{\rm A}$ denotes the 
Alfv\'enic Mach number in the shock reference frame ($M_{\rm A}=\upsilon_{\rm s}/\upsilon_{\rm A}$, where $\upsilon_{\rm A}$ is the Alfv\'en velocity). Simple algebraic manipulation gives the energy density of non-resonantly amplified magnetic field:

\begin{equation}
\epsilon^{(1)}_{\rm B}\approx \frac{1}{2} \frac{\sigma_{\rm cr}}{c} \rho_0 \upsilon_{\rm s}^3,
\end{equation}

\noindent and subsequently, for resonant modes

\begin{equation}
\label{eq:resonant}
\epsilon^{(2)}_{\rm B}\approx  \frac{3}{2} \frac{\sigma_{\rm cr}}{\tilde{M_{\rm A}}}
\rho_0 \upsilon_{\rm s}^2.
\end{equation}

\noindent We can then obtain the ratio between energy densities of non-resonantly and resonantly amplified magnetic fields:

\begin{equation}
\lambda \approx \frac{1}{3} \frac{\upsilon_{\rm s}}{c} \tilde{M_{\rm A}},
\end{equation}

\noindent Therefore, we introduce a correction $(1+\lambda)$ to the original relation for resonant MFA \citep{caprioli09}, in order to account for resonant and non-resonant streaming instabilities:
 
 \begin{equation}
\label{eq:Pw-res}
\frac{P_{\mathrm{w},p}}{\rho_{0}\upsilon_{\rm s}^{2}} \cong \frac{1-\zeta}{4M_{A,0}} U_p^{-3/2}(1-U_p^{2}) (1+\lambda).
\end{equation}

\noindent Here, $P_{\mathrm{w},p}$ denotes precursor magnetic pressure of Alfv\'en waves  
at point "$p$" in the precursor \citep[see e.g.][]{blasi04}, $U_p$ represents the dimensionless fluid velocity $\upsilon_p/\upsilon_{\rm s}$ and $\zeta$ is the Alfv\'en wave dissipation parameter (for details, see P17 and references therein).
The ratio $\lambda$ tends to zero as the SNR approaches the later Sedov phase and therefore, Equation (\ref{eq:Pw-res}) reduces to the equation previously used in P17, where resonant MFA dominates.

\subsection{Theoretical background and expectations}
\label{subsec:theory}

Our purpose here is to apply a simplified analytical approach in order to predict some results of our simulations, which can be later used for verification. We analyze the behavior of radio surface brightness in the Sedov phase of evolution in which most SNRs spend the largest part of their lives.

\begin{figure}
\plotone{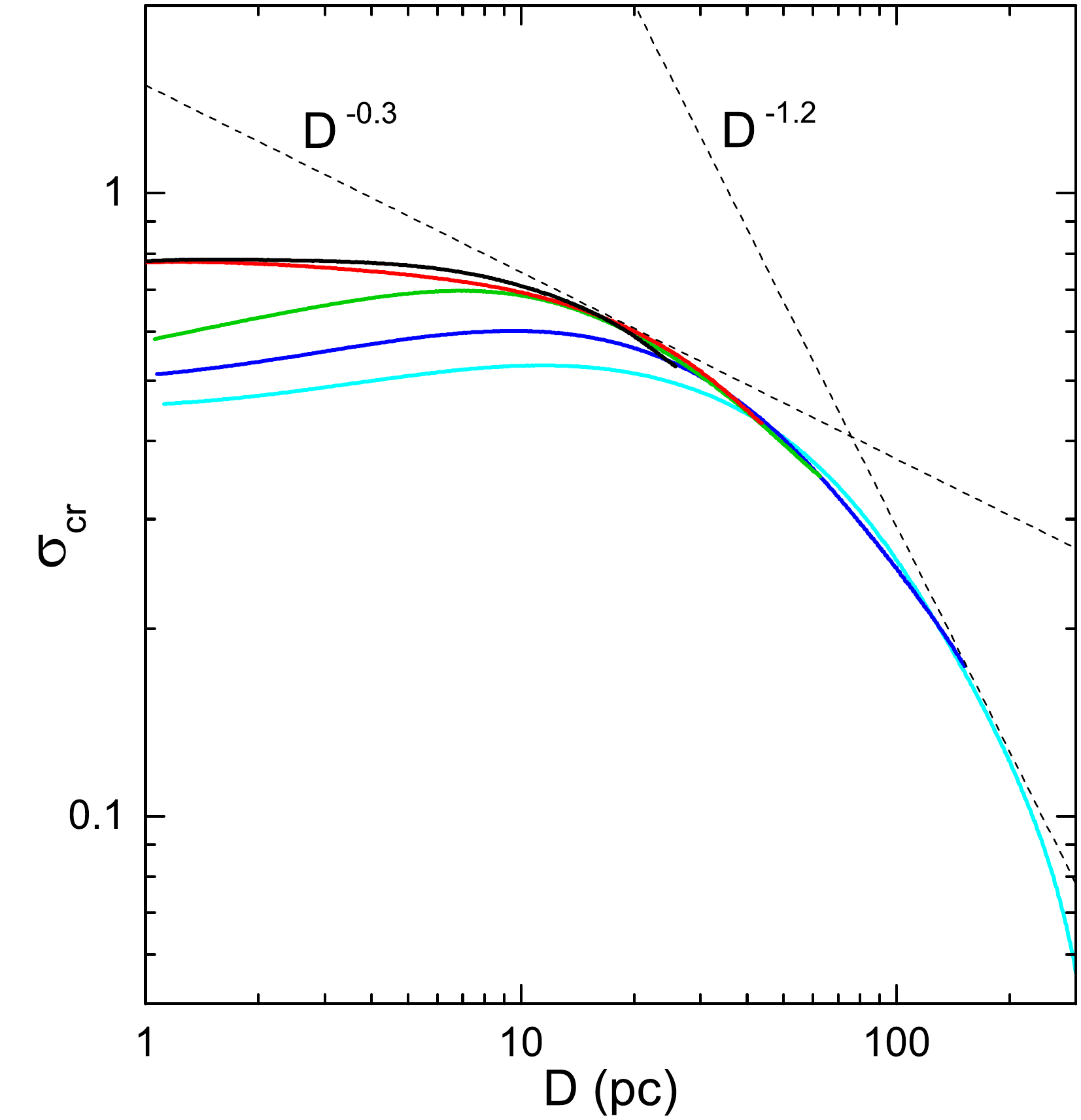} 
\caption{Evolution of $\sigma_{\rm cr}$ ratio, representing the CR pressure at the shock normalized to the shock ram pressure $\rho_0 \upsilon_{\rm s}^2$. Different line colors correspond to different ambient densities, namely $n_{\rm H}/\rm{cm}^{3}=$ 0.005 (cyan), 0.02 (blue), 0.2 (green), 0.5 (red) and 2.0 (black).
\label{fig:Sigma_CR}}
\end{figure}

Total CR energy density is, assuming a power-law
momentum distribution and neglecting energy losses, approximately 
\citep{arbo12}:

\begin{equation}
\epsilon _{\mathrm{CR}} =  K_e (m_e c^2)^{2-\gamma}\frac{\Gamma
(\frac{3-\gamma}{2})\Gamma (\frac{\gamma -2}{2})
}{2\sqrt{\pi}(\gamma -1)} (1+\kappa),
\label{e_CR}
\end{equation}

\noindent $\kappa$ represents the energy ratio between ions and electrons,  $\gamma$ is the energy spectral index ($2 < \gamma < 3$) and $K_e$ is the constant in the power-law energy distributions for the electrons 
$N(E)dE = K_{\rm{e}}E^{-\gamma}dE$.

The radio flux density
of synchrotron radiation of ultra-relativistic electrons, obtained from \citet{pach70} after substituting the emission coefficient  $\varepsilon_{\nu}$ with flux density $S_{\nu}$, is

\begin{equation}
S_{\nu}  \propto K_{\rm{e}} B^{1+\alpha} V \nu^{-\alpha} \;\;\; \rm{\frac{W}{{m}^{2} \, Hz}},
\label{flux-dens}
\end{equation}

\noindent where $B$ is the magnetic field strength, $V$ is the volume, $\nu$ is the frequency and $\alpha$ is the synchrotron spectral index
defined as $\alpha = (\gamma-1)/2$.
Then, radio surface brightness, defined as $\Sigma_{\nu} = S_{\nu} /  \Omega$ where $\Omega$ (in steradians) is the solid
angle of the radio source, scales as:
\begin{equation}
\Sigma_{\nu} \propto S_{\nu} D^{-2} = K_{\rm{e}} B^{1+\alpha} D \;\;\;  \rm{\frac{W}{{m}^{2} \, Hz \, sr}}.
\label{eq:Sigma_D}
\end{equation}

\noindent From Equation~\ref{e_CR} we can deduce:

\begin{equation}
 K_e \propto \epsilon _{\mathrm{CR}} \propto \sigma_{\rm cr} \rho_0 \upsilon_{\rm s}^2,
\label{eq:e_CR}
\end{equation}

\noindent where we don't use equality on the right side because the fraction of the shock energy density in cosmic rays (ions + electrons) differs from $\sigma_{\rm cr}$ (defined in Section~\ref{subsec:mfa}) up to a factor $(\gamma_{\rm cr}-1)$, where 
$\gamma_{\rm cr} \simeq 4/3$ is the adiabatic index of the particles "fluid".

As we already pointed out in Section~\ref{subsec:mfa}, resonant modes dominate in the Sedov phase (Equation~\ref{eq:resonant}) and therefore,
we have:

\begin{equation}
B \propto (\sigma_{\rm cr} \upsilon_{\rm s})^\frac{1}{2}.
\label{eq:B_sedov}
\end{equation}

Evolution of diameter in Sedov phase can be described with $D \propto t^\frac{2}{5}$ \citep{sedov59} and this leads to $\upsilon_{\rm{s}} \propto t^{-\frac{3}{5}} \propto D^{-\frac{3}{2}}$. Substituting Equations~\ref{eq:e_CR} and \ref{eq:B_sedov} into Equation~\ref{eq:Sigma_D} leads to:

\begin{equation}
\Sigma_{\nu}  \propto \sigma_{\rm cr}^\frac{\alpha+3}{2} D^{-\frac{3\alpha+11}{4}},
\label{Sigma_der}
\end{equation}

\noindent Substituting an average SNR spectral index $\alpha=0.5$, reduces to $\Sigma_{\nu} \propto \sigma_{\rm cr}^{1.75} D^{-3.125}$.

Figure~\ref{fig:Sigma_CR} shows the evolution of $\sigma_{\rm cr}$ ratio versus SNR diameter, extracted from simulations with initial energy of $10^{51}$ erg. \noindent We conclude therefore that evolution of $\sigma_{\rm cr}$ should be approximated by $\sigma_{\rm cr}^\frac{\alpha+3}{2} \approx D^{-0.5}$ contribution in earlier and $D^{-2}$ in the later stages (corresponding to limit cases $\sigma_{\rm cr} \propto D^{-0.3}$ and $\sigma_{\rm cr} \propto D^{-1.2}$, see Figure~\ref{fig:Sigma_CR}). The simplified theoretical approach, together with limited insights from NLDSA modeling, predicts radio evolution roughly between $\Sigma_{\nu} \propto D^{-3.5}$ and $\Sigma_{\nu} \propto D^{-5}$, even for spectral slope $\alpha=0.5$, expected in test-particle regime\footnote{Accelerated particles are treated as test particles, having no dynamical role.}. This consideration, however, gives expected dependence for a limited period in evolution. We expect our numerical simulations to give precise insights into broad temporal evolution and contributions of different physical parameters.

\section*{}
\section{Results} 
\label{sec:results}

\begin{deluxetable*}{cccccc}
\tablenum{1}
\tablecaption{Adopted parameters and initial conditions for the hydrodynamic models used to obtain radio evolution of different SNRs
\label{tab:Param}}
\tablewidth{0pt}
\tablehead{
\colhead{Model} & \colhead{Ejecta} & \colhead{Explosion} & \colhead{Ambient} & \colhead{Maximum} & \colhead{Maximum size}\\
\colhead{abreviation} & \colhead{mass} & \colhead{energy} & \colhead{density} & \colhead{age} & \colhead{of physical grid}\\
\colhead{} & \colhead{(M$_{\odot}$)} & \colhead{(10$^{51}$ erg)} & \colhead{(cm$^{-3}$)} & \colhead{(kyr)} & \colhead{(pc)} } 
\decimalcolnumbers
\startdata
SNR0.005\_0.5 & 1.4  & 0.5 & 0.005 &  400 & 140\\
SNR0.005\_1.0 & 1.4 & 1.0 & 0.005  & 400 & 160\\
SNR0.005\_2.0 & 1.4 & 2.0 & 0.005 & 500 & 200\\
SNR0.02\_0.5 & 1.4 & 0.5 & 0.02 &   150 & 80\\
SNR0.02\_1.0 & 1.4 & 1.0 & 0.02 & 150 & 80\\
SNR0.02\_2.0 & 1.4 & 2.0 & 0.02 & 150 & 90\\
SNR0.2\_0.5 & 1.4 & 0.5 & 0.2 &  60 & 35 \\
SNR0.2\_1.0 & 1.4 & 1.0 & 0.2 & 60 & 35\\
SNR0.2\_2.0 & 1.4 & 2.0 & 0.2 & 70 & 35\\
SNR0.5\_0.5 & 10 & 0.5 & 0.5 & 35 & 20\\ 
SNR0.5\_1.0 & 10 & 1.0 & 0.5 & 40 & 25\\
SNR0.5\_2.0 & 10 & 2.0 & 0.5 & 50 & 32\\
SNR2.0\_0.5 & 10 & 0.5 & 2.0 & 23 & 20\\
SNR2.0\_1.0 & 10 & 1.0 & 2.0 & 23 & 20\\
SNR2.0\_2.0 & 10 & 2.0 & 2.0 & 23 & 20\\
\enddata
\end{deluxetable*}

\begin{figure*}
\plotone{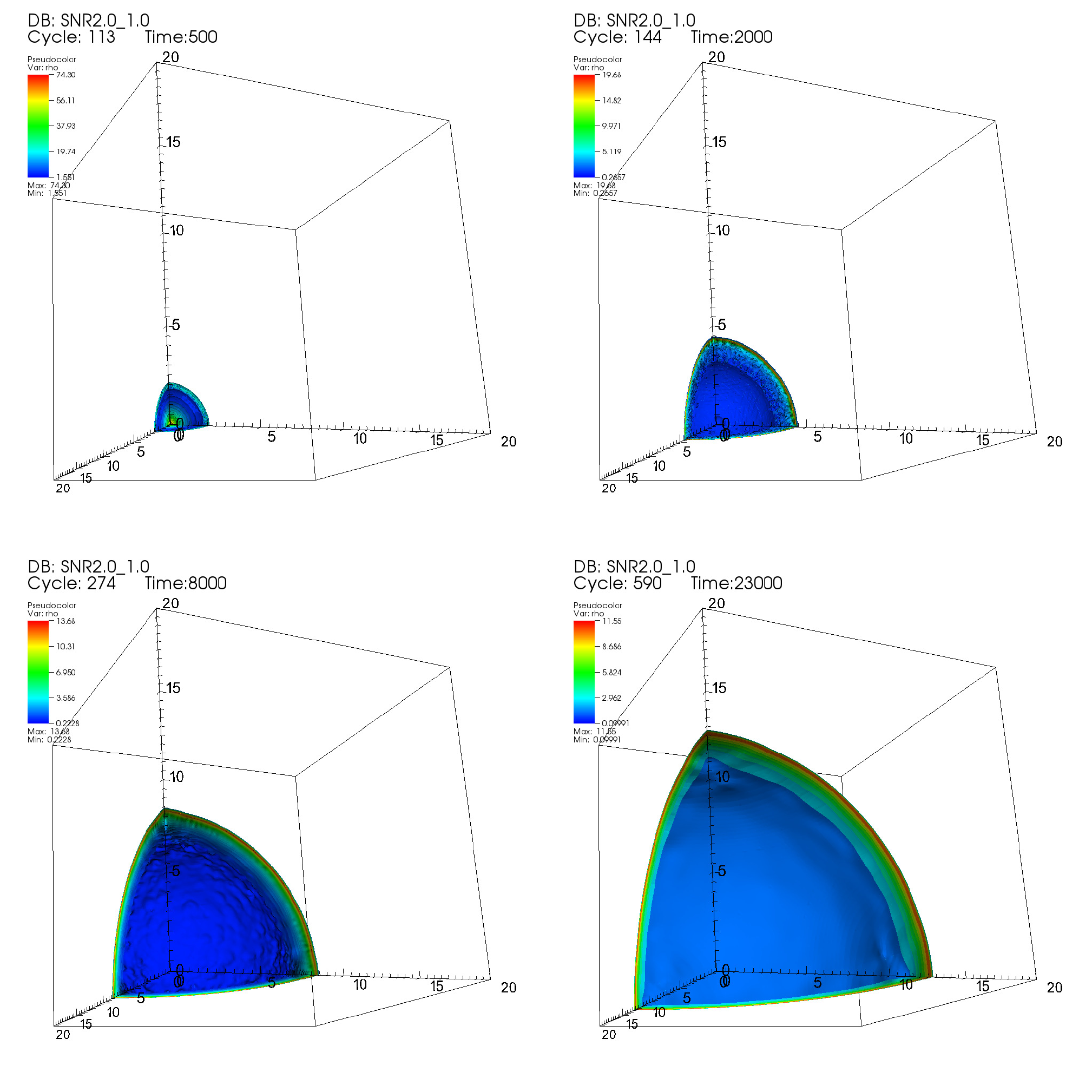} 
\caption{The figure shows the simulation domain (SN explosion occurred at the origin of the one octant in 3D Cartesian coordinate system $(x_0, y_0, z_0)=(0,0,0)$) and the colored regions mark particle number density.
Series of density isosurfaces (3D surface containing cells with the same density value) depict hydrodynamic evolution corresponding to model SNR2.0\_1.0 (see Table~\ref{tab:Param}) at times: $t=500$~yr (upper left panel), $t=2000$~yr (upper right), $t=8000$~yr (lower left) and $t=23000$~yr (lower right). Contours correspond to linearly scaled values between the lowest and the highest values sampled in the intershock region. The box is 20 pc along each axis. Effective AMR resolution varied from $8192^3$ initially to $512^3$ at the end of simulation (23000 yr).
\label{fig:3Dhydro}}
\end{figure*}

\begin{figure*}
\plotone{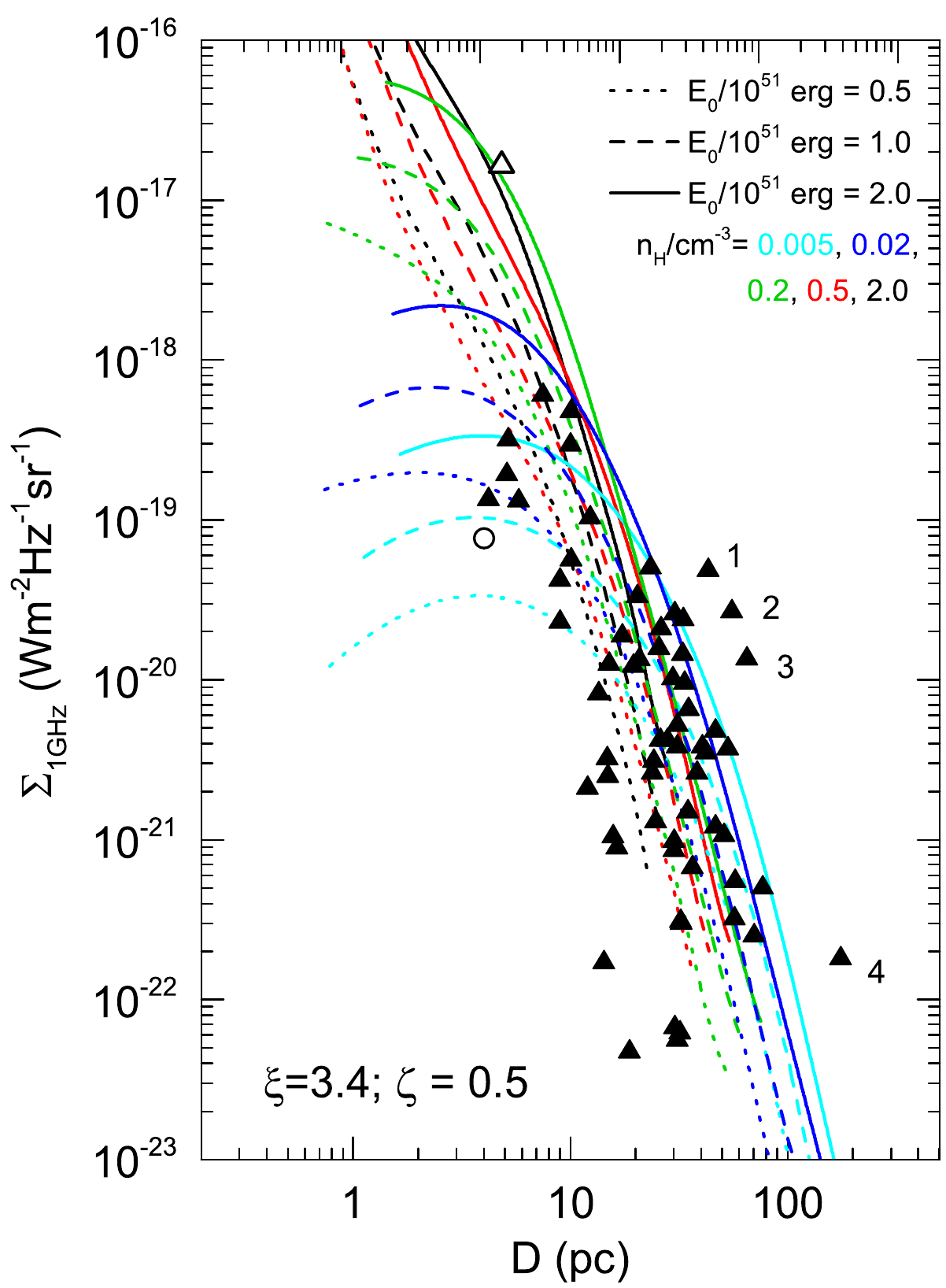}
\caption{Radio surface brightness to diameter diagram for SNRs at frequency 
$\nu$ = 1 GHz, obtained from our numerical simulations. Different line colors correspond to different ambient densities, namely $n_{\rm H}/\rm{cm}^{3}=$ 0.005 (cyan), 0.02 (blue), 0.2 (green), 0.5 (red) and 2.0 (black). Different line styles correspond to the different explosion energies, $E_{\rm 0}/10^{51} \rm{erg}$ = 0.5 (dotted), 1.0 (dashed) and 2.0 (solid). Experimental data represent 65 Galactic SNRs with known distances (triangles) taken from \citet{pavlovic14}. Cassiopeia A is shown with an open triangle while an empty circle represents the youngest Galactic SNR G1.9+0.3 (see P17 for detailed modeling). Numbers represent the following SNRs: (1) CTB 37A, (2) Kes 97, (3) CTB 37B and (4) G65.1+0.6. We show evolutionary tracks for representative case with injection parameter $\xi=3.4$ and non-linear magnetic field damping parameter $\zeta=0.5$.
\label{fig:MW-traka}}
\end{figure*}

\begin{figure*}
\epsscale{1.2}
\plotone{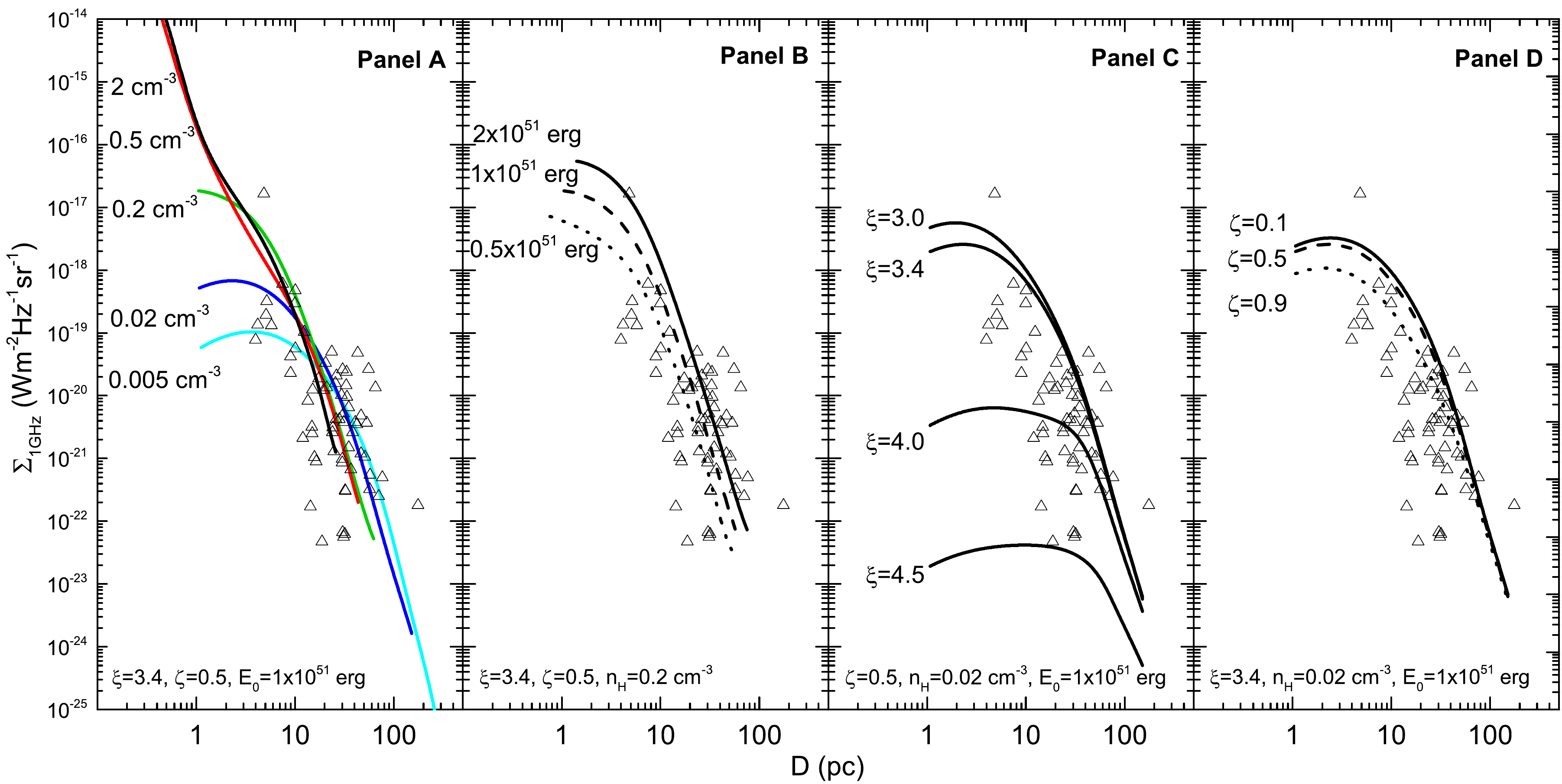}  
\caption{Influence of different simulation parameters on the nature of SNR radio evolutionary tracks. We present here four panels and each of them shows radio evolution in case that we keep all but one parameters fixed. We explore the dependence on the number density of the ambient environment $n_{\rm H}$ (Panel A), the explosion energy $E_0$ (panel B), the CR injection parameter $\xi$ (panel C) and non-linear magnetic field damping parameter $\zeta$ (Panel D). In the lower left corner of each panel we give the values of fixed parameters, chosen as a representative cases. 
\label{fig:MW-2traka}}
\end{figure*}

\begin{figure*}
\plotone{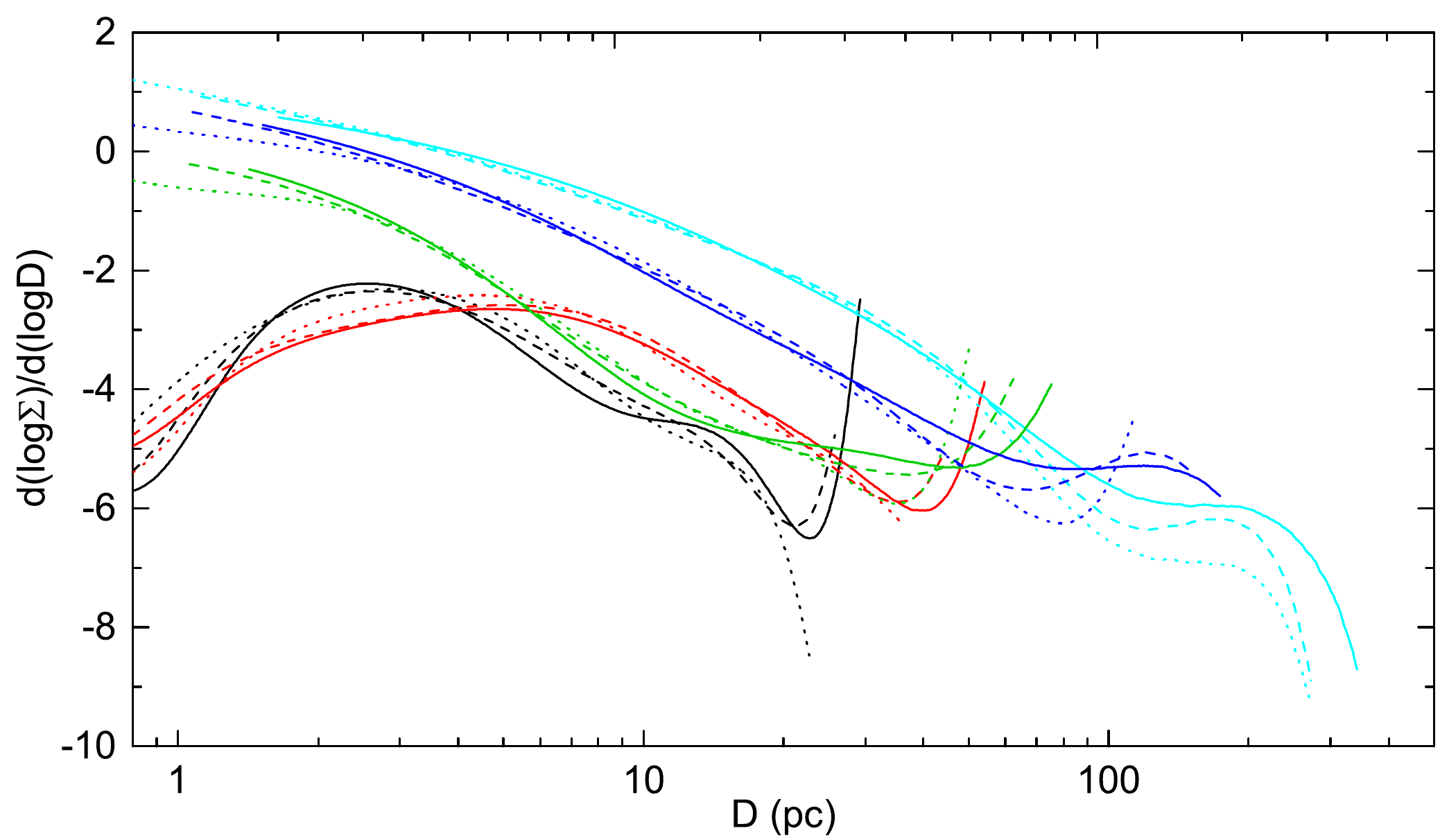} 
\caption{Numerical logarithmic derivative of radio surface brightness with respect to diameter $\frac{\rm{d}\log{\Sigma_{\nu}}}{\rm{d}\log{D}}$. 
Different styles and colors of lines correspond to the same cases as in Figure~\ref{fig:MW-traka}. Evolutionary tracks shown have injection parameter $\xi=3.4$ and magnetic field damping parameter $\zeta=0.5$.
\label{fig:nagib}}
\end{figure*}

We performed our 3D HD simulations (Figure~\ref{fig:3Dhydro}) describing the expansion of SNRs in Cartesian coordinates with the PLUTO code. We adopted the model described in P17, with an improved treatment of MFA. Along with hydrodynamic evolution our code calculates the particle distribution and corresponding synchrotron radio emission from the SNR at any given age.

The purpose of the paper is not to model particular SNRs with the entire set of observable dynamical and spectral characteristics. We rather use a confined set of representative parameters and see if we are able to fit the observational data in a satisfactory way. Our simulations should be appropriate for the observed population, even though we cannot expect precise results for each individual object separately, as these objects will naturally have differences.

The CR injection momentum parameter $\xi$ can typically be in the range 3.0$-$4.5,
where high values of $\xi \ge 4.0$ correspond almost to the test-particle regime 
 and low values of $\xi \le 3.5$ imply efficient DSA \citep{kosenko14}. 
 We adopt the common value $\xi = 3.4$, but also run simulations with $\xi = 3.3$ and
 $\xi = 3.2$ in order to study the sensitivity of the calculations
to the value of this parameter.

Parameter $\zeta$ determines the amount of energy 
in the MHD waves that is dissipated as heat in the plasma
through nonlinear damping processes. Some damping is likely and we
arbitrarily set it to median value $\zeta=0.5$, as a reasonable estimate
\citep[see, for example][]{kang13, ferrand14}.

In our simulations we use proton-to-electron ratio $K_{\rm ep} = 10^{-2}$, as observed in the local CR spectrum and it seems to be characteristic for the later stages (e.g. Sedov) of SNR evolution \citep{sarb17}. On the other hand, this could result in overestimating the radio emission from young SNRs (this will actually turn out to be the case for G1.9+0.3\footnote{In paper P17 we obtained $K_{\rm ep} = 2 \times 10^{-3}$.}).

As indicated by \citet{bif04}, injection takes place only at some fraction of the shock surface, depending on the size of the SNR. This means that radio flux in a spherically symmetric model must be renormalized i.e. reduced by some factor which can vary from case to case. We chose to omit this kind of reduction in order to obtain the upper limit of the simulated evolutionary tracks. 

Table~\ref{tab:Param} summarizes the hydrodynamic parameters adopted. In 
Figure~\ref{fig:MW-traka} we present the simulated radio surface brightness\footnote{It is expressed in units of $\rm{Wm}^{-2}\rm{Hz}^{-1}\rm{sr}^{-1}$ and independent of the distance to the source, as long as the effects of diffraction and extinction can be neglected \citep{radio13}.} $\Sigma_{\nu}$, at frequency $\nu = 1~\rm{GHz}$ as a function of SNR diameter $D$. The data points overplotted in the Figure~\ref{fig:MW-traka} represent the observations, containing 65 Galactic shell SNRs (including Cassiopeia A) with known distances \citep{pavlovic14} and additionally, the youngest Galactic SNR G1.9+0.3.

The simulated dependence of SNR radio surface brightness evolution with the diameter (Figure~\ref{fig:MW-traka} ), calculated for typical hydrodynamic parameters given in Table~\ref{tab:Param}, covers the region of the Galactic experimental points in a very satisfactory way. There are four prominent SNRs: CTB 37A, Kes 67, CTB 37B and G65.1+0.6 (marked with numbers 1 to 4, Figure~\ref{fig:MW-traka}), having significantly higher radio surface brightnesses than expected from our models. This is, however, not surprising as observations suggest that all of them are interacting with molecular clouds, hence explaining the high radio surface brightness.
In the case of SNR CTB 37A (G348.5+0.1), SNR shock-interactions with molecular clouds implied by the presence of 1720\,MHz OH maser emission  \citep{frail96} towards very broad molecular components \citep{reynoso00} which also contain dense ($>10^3$\,cm$^{-1}$) clumps \citep{nigel13}.
Similar applies to the SNR CTB 37B (348.7+0.3) as it resides one of the most active regions in our Galaxy, where a number of shell structures is probably associated with recent SNRs \citep{kassim91} and OH maser sources are detected in the radio band \citep{frail96}. It has been suggested by \citet{dubner99}, \citet{dubner04}, \citet{tian07} and \citet{paron12} that Kes 67 (G18.8+0.3) is interacting with dense molecular gas. \citet{froe15} put G65.1+0.6 on their list of SNRs with identified extended H$_2$ emission line features in their survey. They propose that a possible interaction with a coincident molecular cloud makes G65.1+0.6 a prime target for TeV gamma-ray observations. Type Ia SNRs evolve through low-density media and do not experience severe deceleration. Therefore, encountering dense molecular clouds while still having a quite high Mach number (around a few hundred) makes appropriate conditions for efficient CR acceleration and enhanced radio emission.

Explanation for the observations which lie below modeled tracks is less a demanding task. We recall that simulations were carried-out with the assumption that injection takes place on the entire shock surface. If it takes place only on some fraction, total radio emission will be lower.
Also, it can be inferred from Figure~\ref{fig:MW-2traka} (Panel C) that
injection parameters higher than $\xi=3.4$ causes a significantly lower fraction $\eta$ of the particles to be "injected" in the acceleration process. This directly leads to lower global radio emission. Nevertheless, 
it remains unclear what could cause such an inefficient injection in particular SNRs.

During the earlier SNR evolution, roughly up to around 10 pc, the surface brightness shows relatively high sensitivity to the values of the explosion energy $E_{0}$, the ambient gas number density $n_{\rm H}$, thermal injection parameter $\xi$ and also Alfv\'en heating parameter $\zeta$ (see Figure~\ref{fig:MW-2traka} showing four panels in which the models explore the dependence on any of the above mentioned parameters). It seems from Panel C (Figure~\ref{fig:MW-2traka}) that models show particularly pronounced dependence on injection $\xi$. However, have in mind that we intentionally cover a very broad range of $\xi$ values, corresponding to roughly five orders of magnitude for the ratio $\eta$ of particles injected in the acceleration \citep[$\eta \sim 10^{-2}$ to $10^{-7}$, for subshock compression around 4, see e.g.][]{blasi05}.

Figure~\ref{fig:MW-2traka} also indicates that radio evolutionary tracks of smaller SNRs are more dependent on the variations in basic simulation parameters. In the later evolution these dependencies weaken and
evolutionary tracks tend to cover a relatively narrow region.

Evolutionary tracks for type Ia SNRs evolving in lower density media\footnote{In case of Type Ia events, this is generally fulfilled and we expect interaction with undisturbed, low-density ISM \citep{reyn08}. However, some SNRs like Kepler, N103B and possibly  3C 397 evolve in quite an inhomogeneous environment.} reach maximum radio surface brightness for relatively small diameters (order of few parsecs) and then follow a declining trend. Generally, the diameter which corresponds to the maximum surface brightness increases with decreasing ISM density. Evolutionary tracks corresponding to $n_{\rm H}=0.2$ cm$^{-2}$ show a declining trend during the entire life of SNR. This is not in contradiction with conclusions on radio flux density $S_{\nu}$ evolution derived in P17, but simply a consequence of the relation between the two quantities $\Sigma_{\nu} \propto S_{\nu}D^{-2}$. The radio evolution for CC SNRs complements the trend obtained for type Ia SNRs and their radio evolutionary tracks do not contain a 'brightening phase', therefore representing a monotonically decreasing function of SNR diameter because of the initial interaction with the CSM i.e. stellar wind.

When a SNR approaches the end of the Sedov phase, the CR acceleration
efficiency also decreases as a result of the gradually decreasing Alfv\'en Mach number $M_{\rm A}$. Figures~\ref{fig:MW-traka} and~\ref{fig:MW-2traka} clearly show that acceleration efficiency does not significantly influence radio surface brightness evolution for SNRs in this phase. Also, it has been suggested that higher surrounding ISM density necessarily leads to the greater synchrotron emission from the SNR \citep[see, e.g.][]{duric86, arbutina05}. Figure~\ref{fig:MW-traka} demonstrates that evolutionary tracks of SNRs in dense environments are not necessarily above those residing in lower-density interstellar media, especially for later phases where this conclusion should have the utmost importance.
Denser environment lead to a significant slowdown of the shock wave
and therefore, less efficient acceleration of particles.
According to an analytical theory, based on Bell's test-particle DSA, radio continuum surface brightness of an SNR should scale as $\Sigma_{\nu} \propto B^{1+\alpha} n_{\rm H} D^{-2}$, for a given shock velocity, where $\alpha$ is the synchrotron spectral index \citep{bell78b,duric86}. Therefore, it was intuitively expected for radio evolutionary tracks for SNRs in dense ambient media to lie above those corresponding to SNRs in low density media. This was also one of the starting theoretical assumptions for the study of radio evolution of SNRs, hence making their classification based on ambient density \citep{arbutina05}. However, our simulations show that this is not such a clear trend for SNRs with diameters of a few tens to a few hundred parsecs. We found that SNRs in lower density media show higher radio surface brightness in comparison with those evolving in denser ISM, for a given diameter. Although it may seem counterintuitive, this is actually expected if an accurate treatment of hydrodynamical evolution is performed. Evidently, the forward shock of SNR encountering denser material decelerates more rapidly, sometimes leading to nearly 10 times lower sonic Mach number than those in low density media (for the same corresponding diameter). Higher sonic Mach number means higher injection energy and a higher energy gain during recrossing from upstream to downstream and vice versa. Such a difference will result in а higher number of electrons accelerated to $\sim$ GeV energies (mainly responsible for production of radiation by the synchrotron mechanism) in low density media and therefore higher radio surface brightness. Our simulations imply that any classification of SNRs, based on ambient density and their position on radio surface brightness evolutionary diagram may sometimes be ambiguous and requires caution. This is not the case for smaller diameters (younger SNRs) as the difference between Mach numbers is not so pronounced and also injection energy is relatively high due to high downstream temperature.

Traditionally, statistical studies \citep[see e.g.][and references therein]{urosevic02,pavlovic13,pavlovic14} often proposed the dependence $\Sigma_{\nu}=AD^{\beta}$, based on physical arguments, and used an observational sample to derive parameters $\beta$ and $A$. However, any single relation would represent only an averaged evolutionary track for a sample of SNRs. Slope parameter $\beta$ can be seen as a quantitative description of SNR radio surface brightness evolution with respect to diameter. We also extract $\beta$ evolution from our simulations (Figure~\ref{fig:nagib}) by simple numerical calculation of $\frac{\rm{d}\log{\Sigma_{\nu}}}{\rm{d}\log{D}}$ and by applying Savitzky-Golay\footnote{The Savitzky-Golay method performs a polynomial regression to the data points in the moving window \citep{savitsky64}.} smoothing filter. We conclude from Figure~\ref{fig:nagib} that the evolution of the $\Sigma_{\nu}-D$ slope depends on ambient density and much less on the explosion energy. For more evolved SNRs, namely those having diameters between 10 and few 100 parsecs (Sedov phase), $\beta$ approaches the values approximately between $-6$ and $-4$. The value for empirical $\beta$ obtained in \citet{pavlovic13, pavlovic14} is $\approx-5$ for a Galactic sample. However, one must have in mind that slopes in \citet{pavlovic13, pavlovic14} were obtained by applying a fitting procedure to the entire sample. It is hard to distinguish evolutionary phases of the radio surface brightness in our simulations and connect them to the corresponding phases in SNR evolution, as done semi-analytically by \citet{bif04}. One of the reasons  probably lies in their simplified approach used for the description of SNR dynamics which treats the ejecta as initially expanding as a whole with the single speed $V_0$.

Interestingly, \citet{kostic16} concluded, by using a statistical approach together with fractal density structure of the ISM, that the slope of the surface-brightness evolution steepens if the ambient density is higher. Our simulations partly support this conclusion. Namely, for smaller SNRs ($D \lesssim 20~\rm{pc}$; see Figure~\ref{fig:nagib}), whereas for larger diameters these slopes tend to have density-independent values
between $-6$ and $-4$.

We can conclude from our simulations that the spread in SNR surface brightnesses at a given SNR diameter $D$ is not only due to the spread of the explosion energy $E_0$, but also due to ambient density. Also, we should keep in mind that our simulations don't apply renormalization accounting for injection taking place only on some fraction of the shock surface. This parameter can also produce additional scatter on $\Sigma-D$ diagram. Evolutionary tracks tend to be parallel and form approximately regular shapes of a reasonable width for diameters greater than
$D \sim 10$ pc. This may be seen as the theoretical basis for the
$\Sigma-D$ diagram as an instrument for the distance determination
to SNRs. However, measuring the horizontal width of the region bounded by simulated evolutionary curves, for surface brightnesses $10^{-20}, 10^{-21}$ and $10^{-22}~\rm{W} \rm{m}^{-2} \rm{Hz}^{-1} \rm{sr}^{-1}$ gives a typical error of $\approx$50\% for the calculated lower limit of SNR diameter (distance).

Figure~\ref{fig:MW-traka} also demonstrates that exists a smooth transition between evolutionary tracks of two types of SNRs, those originating from type Ia and others from CC SNe (type Ia SNRs in dense medium are pretty close to those originating from CC events). This makes the eventual determination of the exact SN type of a SNR progenitor only from radio data impossible, implying a requirement for more detailed multi-wavelength observations.

\begin{figure*}
\plotone{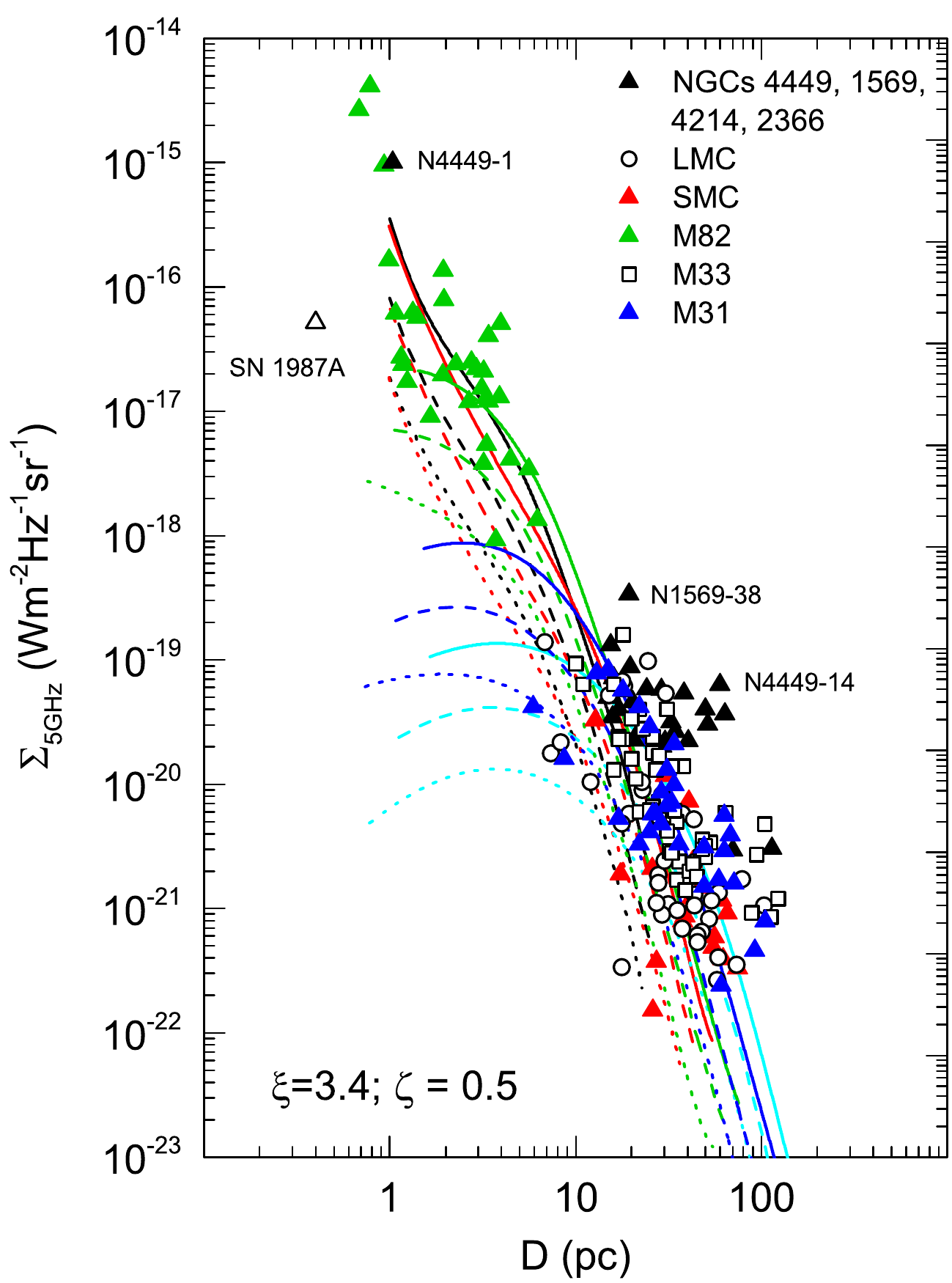}
\caption{Radio surface brightness to diameter diagram for SNRs, at frequency $\nu$ = 5 GHz, obtained from our numerical simulations. The overplotted  SNRs represent the observed samples from the following galaxies: M82 (green); NGC 4449, NGC 1569, NGC 4214, NGC 2366 (black); M31 (blue); SMC (red);  M33 (open squares); LMC (open circles). Although belonging to LMC, we distinguish very young SNR 1987A (open triangle), originating from  the closest SN explosion seen in the modern era.
Different styles and colors of lines correspond to the same cases as in Figure~\ref{fig:MW-traka}.
\label{fig:Extragalactic}}
\end{figure*}

In Figure~\ref{fig:Extragalactic} we present the simulated radio surface brightness at frequency $\nu=5$~GHz, in order to check if it fits the extragalactic samples of SNRs as well. The samples used here, containing available extragalactic SNR populations, were mostly extracted from \citet{bozz17}. Additional samples such as NGC 6744 should be included in future \citep{yew18}. We exclude the Arp 220 sample because it consists of SNRs with diameters below the initial diameter for the expanding ejecta in our simulations. Therefore, Figure~\ref{fig:Extragalactic} contains 215 SNRs in total, overplotted along with the  modeled 5 GHz radio evolutionary tracks. The observational sample contains the remnant of SN 1987A for illustrative purposes, while its complex morphology requires more advanced and specialized treatment \citep{orlando15}. We also find a good agreement of observations with the numerical results. Significant deviation exists only for the joint sample containing SNRs in four irregular galaxies: NGCs 1569, 2366, 4214 and 4449 \citep[radio fluxes were taken from the survey done by][excluding SNRs with questionable diameters due to VLA's resolution]{chom09a}. In comparisson with radio surface brightness predicted by our models, these galaxies contain SNRs which are atypically luminous, considering their size. The possible explanation may be the high star formation rate (SFR), especially for NGC 1569 and NGC 4449. The brightest SNR in NGC 1569 is N1569-38 and it is only half as luminous as the Cas A.
NGC 4449 contains the very young SNR N4449-1 (also known as J1228+441 or 1AXG J122810+4406), which is extraordinarily luminous, five times more luminous than Cas A. SNR 4449-1's shock wave is likely still interacting with the CSM rather than ISM \citep{biet10}. The second most luminous SNR in this galaxy is N4449-14, with a luminosity 80\% that of Cas A \citep{chom09a}. In a galaxy with higher SFR, we expect a larger population of extremely massive stars, and, if $E_0$ correlates with the mass of the progenitor \citep{chom09b}, this energy can be considerably higher than the maximum in our simulations $E_0=2 \times 10^{51}$ erg.

\section{Discussion}
\label{sec:DisConcl}

We study the time evolution of SNR non-thermal emission in the radio domain, with appropriate treatment of the shocked structure hydrodynamics. The SNR hydrodynamical evolution is computed using the 3D hydrodynamical code PLUTO, on the block-structured AMR grid of variable resolution. We also  account for the time-dependent acceleration of particles at the forward shock and corresponding back-reaction on the fluid, which is computed with Blasi's non-linear semi-analytical model \citep{blasi04,blasi05}.

We have also implemented a model for the amplification of the magnetic field in the vicinity of the SNR shock, to account for CR-driven instabilities. Here we include both resonant and non-resonant modes, for the first time in large scale SNR simulations, by implementing recipes obtained from first-principles, particle-in-cell (PIC) simulations and non-linear magnetohydrodynamic (MHD) simulations of CR-excited turbulence. However, this approach has a higher significance for simulations of young SNRs, while, for older SNRs with lower shock speeds, it reduces to the original equation where resonant modes dominate.

Analytical studies of the aforementioned phenomena often rely on simplified assumptions about the evolutionary stage of SNRs, particle spectra and its evolution, magnetic field evolution, etc. Reliable numerical simulations represent a good way to overcome these limitations, aiming to provide a better understanding of underlying physics and explain the observed statistical properties.

DSA and MFA have been implemented in a number of hydrodynamic codes, where these models implement more or less similar DSA treatments. Impact on hydrodynamics is implemented mainly through effective adiabatic index, which is actually a very approximative approach.  \citet{ellison04} used
an approximate, algebraic model of DSA, containing the essential physics of non-linear acceleration, as described in \citet{bere99} and \citet{ellison00}. Later works, like \citet{lee12}, \citet{ferrand12}  and \citet{orlando16} mostly rely on the static NLDSA calculation developed by 
P. Blasi and coworkers \citep{blasi04,blasi05}. This naturally leads to a very good agreement in the particle spectra obtained in the aforementioned work and ours \citep[see for example, proton and electron spectra in our paper][]{pavlovic13}. We are mainly interested in radio evolution, emitted by non-thermal CRs, therefore
we do not include the thermal population of particle spectra.
We do not include a radio surface brightness profile calculation in our simulations, as we are primarily interested in the integrated radio emission. However, if they were included in our work, we don't expect them to be very different from those obtained by \citet{ferrand12}.

We do not seek to model particular SNR, based on its observable dynamical and spectral characteristics. With a set of representative simulation parameters, we derive some average evolutionary tracks in order to see if  we are able to fit entire, currently available observational data sets in a satisfactory way. We also study the influence of the relevant physical parameters on the SNR radio emission and its evolution. We show that typical hydrodynamic and CR acceleration parameters result in radio evolution consistent with radio observations of Galactic SNRs. Simulations demonstrate that evolutionary tracks of SNRs in dense environments are not necessarily above those evolving in lower-density interstellar media. This is mainly because a denser environment leads to a significant slowdown of the shock wave and therefore, less efficient acceleration of particles. If an SNR evolves in denser environment (also assumed to be homogeneous), this can result in the absence of 'brightening phase' i.e. radio evolution is characterized only by declining surface brightness.

Following the results of this modeling, we additionally consider the 'controversial' usage of the $\Sigma-D$ as a prospective distance determination tool. Evolutionary tracks follow very similar forms for diameters greater than $D \sim 10$ pc. Even in a case of a constant  renormalization parameter for all SNRs (to account for acceleration on some fraction of the shock surface), simulated evolutionary curves will produce an error for diameter (distance) determination of around 50\%. Additional problems also exist due to measurement errors and selection effects \citep[see, e.g.][etc.]{arbutina05, urosevic10, pavlovic14}.
 
We find a good agreement in the $\Sigma-D$ plane between observed SNRs and our numerical results. However, SNRs from galaxies, known to have higher SFR, show a systematic trend above calculated evolutionary tracks. It can be explained with higher explosion energies in denser than average environments due to a larger population of extremely massive stars.

\section{Summary and Conclusions}

We have presented a 3D hydrodynamical modeling of SNRs, also accounting for non-linear 
DSA, MFA and shock modifications. We are mainly studying properties of the radio synchrotron emission of SNRs and its evolution. 

Some of the most essential results of our modeling are the following.

1) We have validated our model on available Galactic and extragalactic observational samples. The simulated dependence of SNR radio evolution is consistent with the range of parameters observed in nature.

2) During the earlier SNR evolution, roughly up to a diameter of around 10 pc, the radio surface brightness shows relatively high sensitivity to the values of the explosion energy, the ambient density, the thermal injection parameter and the Alfv\'en heating parameter. In the later evolution these dependencies weaken.
          
3) Radio evolutionary tracks for SNRs evolving in different ambient densities intersect between $\approx 10$ pc and a few tens of parsecs. $\Sigma-D$ tracks for higher ISM density drop below those corresponding to a low density medium. Therefore, correlating SNR ambient density and position on a $\Sigma-D$ diagram may not always be unambiguous and requires caution.

4) The SNR radio emission light curves may show a decline very early, in cases where SNRs evolve through denser ISM. This can sometimes result in a complete absence of the brightening phase for radio SNRs. The situation may be more complicated for radio SN in CSM-dominated phase.
          
5) SNR shocks leaving rarefied bubbles and encountering dense molecular clouds, while still having quite high Mach numbers (around a few hundred), show enhanced radio emission in comparison with those evolving through dense and homogeneous ISM during the whole SNR evolution.

6) Our simulations give $\Sigma-D$ slopes between -4 and -6 for the full Sedov regime, in good agreement with theoretical expectations and observed samples.

7) If the $\Sigma-D$ relation is to be used as a distance determination tool, simulations show that error could be around 50\%, even if the intrinsic morphological characteristics are neglected.
        
The evolutionary tracks presented here can be very useful for radio observers. They can use them for determination of the evolutionary status for all observationally confirmed Galactic and extragalactic SNRs, for which their age or ambient conditions are unknown.

Additionally, this type of modeling is expected to be a useful apparatus  for future observers working on powerful radio telescopes such as ALMA, 
MWA, ASKAP, SKA and FAST. Large scale surveys should be carefully planed in order to give new discoveries. Having the information about sensitivity limits of the instruments, simulations could help to predict the science output in terms of new detections. Later on, a lot of interesting effects connected with CR acceleration may be detected with high sensitive instruments and having support in high resolution simulations.
\\

The authors would like to thank the anonymous referee for a constructive report and useful comments.
This work is part of project no. 176005 \lq\lq Emission nebulae:
structure and evolution\rq\rq, supported by the Ministry of Education, Science, and Technological Development of the Republic of Serbia. Numerical simulations were run on the PARADOX-IV supercomputing facility at the Scientific Computing Laboratory of the Institute of Physics Belgrade, supported in part by the Ministry of Education, Science and Technological Development of the Republic of Serbia under projects no. ON171017 and OI1611005. S.O. acknowledges support by the PRIN INAF 2014 grant \lq\lq Filling the gap between supernova explosions and their remnants through magnetohydrodynamic modelling and high-performance computing\rq\rq.
M.P. acknowledge the hospitality of the Palermo Astronomical Observatory \lq\lq Giuseppe S. Vaiana\rq\rq~where part of this work was carried out. M.P. is grateful to Gilles Ferrand for discussions, advices and help during this work and coding. PLUTO, the software used in this work, was developed at the Department of Physics of the Turin University in a joint collaboration with INAF, Turin Astronomical Observatory and the SCAI Department of CINECA. M.P. wants to thank Claudio Zanni and Andrea Mignone for their help with the PLUTO code.

\software{PLUTO \citep[Version 4.2; ][]{mignone07,mignone12}, 
Chombo \citep{adams13}}




\begin{thebibliography}{}

\bibitem[Ackermann et al.(2013)]{acker13} Ackermann, M., Ajello, M., Allafort, A., et al.\ 2013, Science, 339, 807 

\bibitem[Adams et al.(2013)]{adams13} Adams, M., Colella, P., Graves, D. T., Johnson, J. N., Keen, N. D. et al., \emph{Chombo Software Package for AMR Applications-Design Document}, 2013,  Lawrence Berkeley National Laboratory Technical Report LBNL-6616E  

\bibitem[Aharonian et al.(2017)]{aharonian17} Aharonian, F., Sun, X.-n., \& Yang, R.-z.\ 2017, \aap, 603, A7 

\bibitem[Amato \& Blasi(2009)]{amato09} Amato, E., \& Blasi, P.\ 2009, \mnras, 392, 1591 

\bibitem[Arbutina \& Uro{\v s}evi{\'c}(2005)]{arbutina05} Arbutina, B., \& Uro{\v s}evi{\'c}, D.\ 2005, \mnras, 360, 76 

\bibitem[Arbutina et al.(2012)]{arbo12} Arbutina, B., Uro{\v s}evi{\'c}, D., Andjeli{\'c}, M.~M., Pavlovi{\'c}, M.~Z., \& Vukoti{\'c}, B.\ 2012, \apj, 746, 79 

\bibitem[Axford et al.(1977)]{axford77} Axford, W.~I., Leer, E., \& Skadron, G.\ 1977, International Cosmic Ray Conference, 11, 132 

\bibitem[Baars et al.(1977)]{baars77} Baars, J.~W.~M., Genzel, R., Pauliny-Toth, I.~I.~K., \& Witzel, A.\ 1977, \aap, 61, 99 

\bibitem[Bandiera \& Petruk(2010)]{bandiera10} Bandiera, R., \& Petruk, O.\ 2010, \aap, 509, A34 

\bibitem[Berkhuijsen(1986)]{berkhujsen86} Berkhuijsen, E.~M.\ 1986, \aap, 166, 257 

\bibitem[Bell(1978a)]{bell78a} Bell, A.~R.\ 1978a, \mnras, 182, 147 

\bibitem[Bell(1978b)]{bell78b} Bell, A.~R.\ 1978b, \mnras, 182, 443

\bibitem[Bell(2004)]{bell04} Bell, A.~R.\ 2004, \mnras, 353, 550 

\bibitem[Berezhko \& Ellison(1999)]{bere99} Berezhko, E.~G., \& Ellison, D.~C.\ 1999, \apj, 526, 385 

\bibitem[Berezhko \& V{\"o}lk(2004)]{bif04} Berezhko, E.~G., \& V{\"o}lk, H.~J.\ 2004, \aap, 427, 525 

\bibitem[Bozzetto et al.(2017)]{bozz17} Bozzetto, L.~M., Filipovi{\'c}, M.~D., Vukoti{\'c}, B., et al.\ 2017, \apjs, 230, 2 

\bibitem[Bietenholz et al.(2010)]{biet10} Bietenholz, M.~F., Bartel, N., Milisavljevic, D., et al.\ 2010, \mnras, 409, 1594 

\bibitem[Blasi(2004)]{blasi04} Blasi, P.\ 2004, Astroparticle Physics, 21, 45 

\bibitem[Blasi et al.(2005)]{blasi05} Blasi, P., Gabici, S., \& Vannoni, G.\ 2005, \mnras, 361, 907 

\bibitem[Blasi(2013)]{blasi13} Blasi, P.\ 2013, \aapr, 21, 70 


\bibitem[Blandford \& Ostriker(1978)]{blandford78} Blandford, R.~D., \& Ostriker, J.~P.\ 1978, \apjl, 221, L29  

\bibitem[Blondin et al.(1998)]{blondin98} Blondin, J.~M., Wright, E.~B., Borkowski, K.~J., \& Reynolds, S.~P.\ 1998, \apj, 500, 342 

\bibitem[Bykov et al.(2014)]{bykov14} Bykov, A.~M., Ellison, D.~C., Osipov, S.~M., \& Vladimirov, A.~E.\ 2014, \apj, 789, 137 

\bibitem[Callingham et al.(2016)]{call16} Callingham, J.~R., Gaensler, B.~M., Zanardo, G., et al.\ 2016, \mnras, 462, 290 

\bibitem[Caprioli et al.(2009)]{caprioli09} Caprioli, D., Blasi, P., Amato, E., \& Vietri, M.\ 2009, \mnras, 395, 895 

\bibitem[Caprioli(2012)]{caprioli12} Caprioli, D.\ 2012, \jcap, 7, 038 

\bibitem[Caprioli \& Spitkovsky(2014)]{capri14} Caprioli, D., \& Spitkovsky, A., 2014, \apj, 794, 46

\bibitem[Chomiuk \& Wilcots(2009a)]{chom09a} Chomiuk, L., \& Wilcots, E.~M.\ 2009a, \aj, 137, 3869

\bibitem[Chomiuk \& Wilcots(2009b)]{chom09b} Chomiuk, L., \& Wilcots, E.~M.\ 2009b, \apj, 703, 370 

\bibitem[Cox \& Anderson(1982)]{cox82} Cox, D.~P., \& Anderson, P.~R.\ 1982, \apj, 253, 268 

\bibitem[De Horta et al.(2014)]{horta14} De Horta, A.~Y., Filipovic, M.~D., 
Crawford, E.~J., et al.\ 2014, Serbian Astronomical Journal, 189, 41

\bibitem[Dubner et al.(1999)]{dubner99} Dubner, G., Giacani, E., Reynoso, E., et al.\ 1999, \aj, 118, 930 

\bibitem[Dubner et al.(2004)]{dubner04} Dubner, G., Giacani, E., Reynoso, E., \& Par{\'o}n, S.\ 2004, \aap, 426, 201 

\bibitem[Duric \& Seaquist(1986)]{duric86} Duric, N., \& Seaquist, E.~R.\ 1986, \apj, 301, 308 

\bibitem[Dwarkadas \& Chevalier(1998)]{dwarka98} Dwarkadas, V.~V., \& Chevalier, R.~A.\ 1998, \apj, 497, 807 

\bibitem[Ellison et al.(2000)]{ellison00} Ellison, D.~C., Berezhko, E.~G., \& Baring, M.~G.\ 2000, \apj, 540, 292 

\bibitem[Ellison et al.(2004)]{ellison04} Ellison, D.~C., Decourchelle, A., \& Ballet, J.\ 2004, \aap, 413, 189 

\bibitem[Fukui et al.(2017)]{fakui17} Fukui, Y., Sano, H., Sato, J., et al.\ 2017, \apj, 850, 71 

\bibitem[Ferrand et al.(2012)]{ferrand12} Ferrand, G., Decourchelle, A., \& Safi-Harb, S.\ 2012, \apj, 760, 34 

\bibitem[Ferrand et al.(2014)]{ferrand14} Ferrand, G., Decourchelle, A., \& Safi-Harb, S.\ 2014, \apj, 789, 49 

\bibitem[Frail et al.(1996)]{frail96} Frail, D.~A., Goss, W.~M., Reynoso, E.~M., et al.\ 1996, \aj, 111, 1651 

\bibitem[Froebrich et al.(2015)]{froe15} Froebrich, D., Makin, S.~V., Davis, C.~J., et al.\ 2015, \mnras, 454, 2586 

\bibitem[Gilfanov \& Bogd{\'a}n(2010)]{gilf10} Gilfanov, M., \& Bogd{\'a}n, {\'A}.\ 2010, \nat, 463, 924 

\bibitem[Green(1991)]{green91} Green, D.~A.\ 1991, \pasp, 103, 209 

\bibitem[Green et al.(2008)]{green08} Green, D.~A., Reynolds, S.~P., 
Borkowski, K.~J., et al.\ 2008, \mnras, 387, L54 

\bibitem[Kang et al.(2013)]{kang13} Kang, H., Jones, T.~W., \& Edmon, P.~P.\ 2013, \apj, 777, 25 

\bibitem[Kassim et al.(1991)]{kassim91} Kassim, N.~E., Weiler, K.~W., \& Baum, S.~A.\ 1991, \apj, 374, 212 

\bibitem[Kne{\v z}evi{\'c} et al.(2017)]{knez17} Kne{\v z}evi{\'c}, S., L{\"a}sker, R., van de Ven, G., et al.\ 2017, \apj, 846, 167 

\bibitem[Koo \& Park(2016)]{handbook16} Koo, B.-C. and Park, C.,
"Supernova Remnant Cassiopeia A", Handbook of Supernovae; (eds. Alsabti, A. W. and Murdin, P.), Astronomy and Astrophysics Library.~Springer-Verlag Berlin Heidelberg, 
2013

\bibitem[Kosenko et al.(2014)]{kosenko14} Kosenko, D., Ferrand, G., \& Decourchelle, A.\ 2014, \mnras, 443, 1390 

\bibitem[Kosti{\'c} et al.(2016)]{kostic16} Kosti{\'c}, P., Vukoti{\'c}, B., Uro{\v s}evi{\'c}, D., Arbutina, B., \& Prodanovi{\'c}, T.\ 2016, \mnras, 461, 1421 

\bibitem[Krymskii(1977)]{krym77} Krymskii, G.~F.\ 1977, Akademiia Nauk SSSR Doklady, 234, 1306 

\bibitem[Lee et al.(2012)]{lee12} Lee, S.-H., Ellison, D.~C., \& Nagataki, S.\ 2012, \apj, 750, 156

\bibitem[Maggi et al.(2016)]{maggi16} Maggi, P., Haberl, F., Kavanagh, P.~J., et al.\ 2016, \aap, 585, A162 

\bibitem[Maxted et al.(2013)]{nigel13} Maxted, N.~I., Rowell, G.~P., Dawson, B.~R., et al.\ 2013, \mnras, 434, 2188 

\bibitem[Mignone et al.(2007)]{mignone07} Mignone, A., Bodo, G., Massaglia, S., et al.\ 2007, \apjs, 170, 228 

\bibitem[Mignone et al.(2012)]{mignone12} Mignone, A., Zanni, C., Tzeferacos, P., et al.\ 2012, \apjs, 198, 7 

\bibitem[Morlino \& Caprioli(2012)]{morlino12} Morlino, G., \& Caprioli, D.\ 2012, \aap, 538, A81

\bibitem[Murphy et al.(2008)]{murphy08} Murphy, T., Gaensler, B.~M., \& Chatterjee, S.\ 2008, \mnras, 389, L23

\bibitem[Nikoli{\'c} et al.(2013)]{nikolic13} Nikoli{\'c}, S., van de Ven, G., Heng, K., et al.\ 2013, Science, 340, 45

\bibitem[Olling et al.(2015)]{olling15} Olling, R.~P., Mushotzky, R., Shaya, E.~J., et al.\ 2015, \nat, 521, 332 

\bibitem[Orlando et al.(2012)]{orlando12} Orlando, S., Bocchino, F., Miceli, M., Petruk, O., \& Pumo, M.~L.\ 2012, \apj, 749, 156 

\bibitem[Orlando et al.(2015)]{orlando15} Orlando, S., Miceli, M., Pumo, M.~L., \& Bocchino, F.\ 2015, \apj, 810, 168 

\bibitem[Orlando et al.(2016)]{orlando16} Orlando, S., Miceli, M., Pumo, M.~L., \& Bocchino, F.\ 2016, \apj, 822, 22 

\bibitem[Pacholczyk(1970)]{pach70} Pacholczyk, A.~G.\ 1970, Series of Books in Astronomy and Astrophysics, San Francisco: Freeman, 1970

\bibitem[Paron et al.(2012)]{paron12} Paron, S., Ortega, M.~E., Petriella, A., et al.\ 2012, \aap, 547, A60 

\bibitem[Pavlovi{\'c} et al.(2013)]{pavlovic13} Pavlovi{\'c}, M.~Z., Uro{\v s}evi{\'c}, D., Vukoti{\'c}, B., Arbutina, B., \& G{\"o}ker, {\"U}.~D.\ 2013, \apjs, 204, 4 

\bibitem[Pavlovi{\'c} et al.(2014)]{pavlovic14} Pavlovi{\'c}, M.~Z., Dobardzic, A., Vukotic, B., \& Urosevic, D.\ 2014, Serbian Astronomical Journal, 189, 25 

\bibitem[Pavlovi{\'c}(2017)]{pavlovic17} Pavlovi{\'c}, M.~Z.\ 2017, \mnras, 468, 1616

\bibitem[Pfrommer et al.(2017)]{pformer17} Pfrommer, C., Pakmor, R., Schaal, K., Simpson, C.~M., \& Springel, V.\ 2017, \mnras, 465, 4500 

\bibitem[Petruk et al.(2016)]{petruk16} Petruk, O., Kuzyo, T., \& Beshley, V.\ 2016, \mnras, 456, 2343 

\bibitem[Reichart \& Stephens(2000)]{raich00} Reichart, D.~E., \& Stephens, A.~W.\ 2000, \apj, 537, 904 

\bibitem[Reynolds(2008)]{reyn08} Reynolds, S.~P.\ 2008, \araa, 46, 89 

\bibitem[Reynolds(2017)]{reyn17} Reynolds, S.~P.\ 2017, arXiv:1708.05386

\bibitem[Reynoso \& Mangum(2000)]{reynoso00} Reynoso, E.~M., \& Mangum, J.~G.\ 2000, \apj, 545, 874 

\bibitem[Sarbadhicary et al.(2017)]{sarb17} Sarbadhicary, S.~K., Badenes, C., Chomiuk, L., Caprioli, D., \& Huizenga, D.\ 2017, \mnras, 464, 2326 

\bibitem[Savitzky \& Golay(1964)] {savitsky64} Savitzky, A., \& Golay, M.~J.~E., 1964, Analytical Chemistry, 36 (8), 1627-1639

\bibitem[Sedov(1959)]{sedov59} Sedov, L.~I.\ 1959, Similarity and Dimensional Methods in Mechanics, New York: Academic Press

\bibitem[Shklovskii(1960a)]{sklovski60a} Shklovskii, I.~S.\ 1960, \azh, 37, 256 

\bibitem[Shklovskii(1960b)]{sklovski60b} Shklovskii, I.~S.\ 1960, \azh, 37, 369 

\bibitem[Slavin et al.(2017)]{slavin17} Slavin, J.~D., Smith, R.~K., Foster, A., et al.\ 2017, \apj, 846, 77 

\bibitem[Tian et al.(2007)]{tian07} Tian, W.~W., Leahy, D.~A., \& Wang, Q.~D.\ 2007, \aap, 474, 541 

\bibitem[Urosevic(2002)]{urosevic02} Urosevic, D.\ 2002, Serbian Astronomical Journal, 165,  27

\bibitem[Uro{\v s}evi{\'c} et al.(2005)]{urosevic05} Uro{\v s}evi{\'c}, D., Pannuti, T.~G., Duric, N., \& Theodorou, A.\ 2005, \aap, 435, 437 

\bibitem[Uro{\v s}evi{\'c} et al.(2010)]{urosevic10} Uro{\v s}evi{\'c}, D., Vukoti{\'c}, B., Arbutina, B., \& Sarevska, M.\ 2010, \apj, 719, 950 

\bibitem[Vukoti{\'c} et al.(2014)]{vukotic14} Vukoti{\'c}, B., Jurkovi{\'c}, M., Uro{\v s}evi{\'c}, D., \& Arbutina, B.\ 2014, \mnras, 440, 2026 

\bibitem[Wilson et al.(2013)]{radio13} Wilson, T.~L., Rohlfs, K., \& H{\"u}ttemeister, S.\, Tools of Radio Astronomy (6th edition); Astronomy and Astrophysics Library.~Springer-Verlag Berlin Heidelberg, 2013

\bibitem[Woods et al.(2017)]{woods17} Woods, T.~E., Ghavamian, P., Badenes, C., \& Gilfanov, M.\ 2017,  Nature Astronomy, 1, arXiv:1709.09190

\bibitem[Yew et al.(2018)]{yew18} Yew, M., Filipovi\' c, M. D., Roper, Q., Collier, J. D., Crawford, E. J. et al., 2018, \pasa~(in preparation)

\bibitem[Zanardo et al.(2010)]{zanardo10} Zanardo, G., Staveley-Smith, L., Ball, L., et al.\ 2010, \apj, 710, 1515




\end{thebibliography}
\end{document}